\documentclass[10pt, conference, compsocconf]{IEEEtran}
\pdfoutput=1
\usepackage[english]{babel}
\usepackage{graphicx}
\usepackage{balance}  
\usepackage{hyperref}

\usepackage{epsfig}
\usepackage{amssymb}
\usepackage{amsmath}
\usepackage[noend]{algorithmic}
\usepackage{algorithm}
\usepackage{multirow}
\usepackage{subfigure}
\usepackage{balance}
\usepackage{bm}

\newtheorem{mydef}{Definition}

\begin{document}


\title{Scalable Locality-Sensitive Hashing for Similarity Search in High-Dimensional, Large-Scale
Multimedia Datasets}

\author{

  \IEEEauthorblockN{Thiago S. F. X. Teixeira} 
  \IEEEauthorblockA{Dept. of Computer Science\\
  University of Brasilia\\
  Brasilia, DF, Brazil \\
  thiagotei@gmail.com} \and 

  \IEEEauthorblockN{George Teodoro}
  \IEEEauthorblockA{Dept. of Computer Science\\
  University of Brasilia\\
  Brasilia, DF, Brazil \\
  glmteodoro@gmail.com} \and 

  \IEEEauthorblockN{Eduardo Valle}
  \IEEEauthorblockA{FEEC\\ 
  State University of Campinas\\
  Campinas, SP, Brazil\\
  dovalle@dca.fee.unicamp.br} \and 

  \IEEEauthorblockN{Joel H. Saltz} 
  \IEEEauthorblockA{Dept. of Biomedical Informatics\\
  Stony Brook University\\
  Stony Brook, NY, USA\\
  joel.saltz@stonybrookmedicine.edu}
\vspace{-14mm}
}

\maketitle

\begin{abstract} 

Similarity search is critical for many database applications, including the
increasingly popular online services for Content-Based Multimedia Retrieval
(CBMR).  These services, which include image search engines, must handle an
overwhelming volume of data, while keeping low  response times. Thus,
scalability is imperative for similarity search in Web-scale applications, but
most existing methods are sequential and target shared-memory machines. Here we
address these issues with a distributed, efficient, and scalable  index based
on Locality-Sensitive Hashing (LSH).  LSH is one of the most efficient and
popular techniques for similarity search, but its poor referential locality
properties has made its implementation a challenging problem. Our solution is
based on a widely asynchronous dataflow parallelization with a number of
optimizations that include a hierarchical parallelization to decouple indexing
and data storage, locality-aware data partition strategies to reduce message
passing, and multi-probing to limit memory usage.  The proposed parallelization
attained an efficiency of $\bm{90\%}$ in a distributed system with
about $\bm{800}$~CPU cores. In particular, the original locality-aware data
partition  reduced the number of messages exchanged in $\bm{30}$\%. Our
parallel LSH was evaluated using the largest public dataset for similarity
search (to the best of our knowledge) with $\bm{10^9}$ $\bm{128}$-d SIFT
descriptors extracted from Web images.  This is two orders of magnitude larger
than datasets that previous LSH parallelizations could handle.

\end{abstract}

%
%


\section{Introduction}
\label{sec:intro}

Similarity search consists in finding in a reference dataset the most similar objects 
to a query object. Multimedia retrieval applications
typically represent objects (images, videos, songs, etc) as
high-dimensional feature vectors. In this context, similarity search is
abstracted as finding the vectors in the dataset which are closest to a query vector, using a given distance (often Euclidean). Similarity search is a core
operation for content-based multimedia retrieval (CBMR) applications such as
image search engines, real-time song identification, tagging of photos in
social networks, etc.~\cite{Bohm:2001:SHS:502807.502809,Jegou2008,5432202,Joly2008a,Valle:2008:HDI:1458082.1458181,teodoro-2013-vldb} Query processing in these
applications may consist of several complex stages, but still similarity search
will be one of the most critical and costly steps.

The success of current Web-scale CBMR applications, such as image search
engines, depends on their ability to efficiently handle very large and
increasing volumes of data, while keeping low the response times as observed by
users. Although the amount of data to be indexed by these applications 
exceeds the capabilities of commodity machines, most of the state-of-the-art
indexing methods have been designed for sequential execution in shared-memory
systems.

In this work, we address the challenges of efficiently performing similarity
search for large-scale CBMR services on distributed environment. We design and
implement a distributed, efficient, and scalable similarity search index based
on Locality-Sensitive Hashing (LSH), 
one of the most efficient and popular
approaches for similarity search~\cite{Lv:2007}. It relies on the use of a
family of locality-preserving hash function, creating several hash tables that
hash together similar objects with high probability. During search phase, the
hash tables can be queried to retrieve a relatively small set of objects which
are good candidates to be the closest to the query object. The distance  is
then computed to those candidates and the best are returned as a (very good)
approximate solution to the similarity search.

LSH indexing has been shown to be a very challenging algorithm for
parallelization in distributed memory machines because of the lack of
referential locality induced by the ``curse of dimensionality''. It is very
difficult to partition the dataset for parallel execution without incurring in
excessive communication during the search phase.  Additionally, on sequential
LSH implementations, efficiency is based upon the use of a large number of hash
tables, each of them indexing the entire dataset.  For shared-memory systems
with uniform access costs, LSH hash tables will typically store references for
the actual data objects, in order to avoid replicating the dataset. This is,
however, an issue distributed environments: since replicating the dataset for
each table is not feasible, once the references are retrieved the actual data
must be accessed and the ``curse of dimensionality'' makes it difficult to
arrange them in order to warrant good locality. If care is not taken, this
phase will drown the system with messages, ruining performance.

The challenges of  parallelizing LSH are addressed in our
work with: (i)~a decomposition of the method into a dataflow of computing
stages, such that the hash buckets storing references to the dataset objects and
the data objects are stored into different application stages in order to avoid data
replication; (ii)~a hierarchical parallelization in which stateful stages of
the application are designed as multi-threaded processes that take advantage of
multiple CPU cores in a node.  It results into a smaller number of data
partitions (one per node instead of one per CPU core), and an improved
scalability because of the reduced communication; (iii)~a study of multiple data
partition strategies to effectively distribute state, which shows that adequate
locality-aware data partition may lead to substantial reduction in
communication; (iv)~a widely asynchronous design that allows for the algorithm
to overlap communication and computation; and, (v)~a multi-probe LSH that allows
for a smaller number of hash tables to be used, while multiple buckets are
visited in each table to achieve the required search quality.

All these propositions have been thoroughly evaluated using what is to the best
of our knowledge the largest publicly available dataset for CBMR applications.
This dataset contains 1~billion 128-dimensional SIFT (Scale-Invariant Feature
Transform)~\cite{Lowe:1999:ORL:850924.851523} feature vectors extracted from images collected in the Web. This
dataset is about two orders of magnitude larger than the largest dataset
processed using LSH indexing methods prior to our work. Additionally, our
parallelization has attained about 90\% of efficiency with the use of 801~CPU
cores/51~nodes. The performance evaluation of the distributed multi-probe
version of LSH attains sublinear increasing in the communication as the number
of probes per table grows, resulting into a surprisingly good trade-off between
performance and search quality. 

The rest of this paper is organized as follows: Section~\ref{sec:background}
introduces the similarity search (nearest neighbors) problem in
high-dimensional spaces and presents some of the most popular indexing methods;
Section~\ref{sec:lsh} details the main concepts used by the LSH indexing as well
as recent improvements in the basic LSH; Our distributed memory parallelization
of the LSH indexing is discussed in Section~\ref{sec:parallelization}; and, the
experimental results and conclusions are presented, respectively, in
Sections~\ref{sec:results} and~\ref{sec:conclusions}.

\section{Background}
\label{sec:background}

\subsection{Nearest Neighbors Search}

The nearest neighbors (NN) search problem has received increasing attention in
the last decades. Several efforts have focused on the development of data
structures, including kd-tree~\cite{Friedman:1977:AFB:355744.355745}, k-means
tree~\cite{Muja09fastapproximate}, cover
trees~\cite{Beygelzimer:2006:CTN:1143844.1143857}, and others that provide a
locality-aware partition of the input data, allowing to prune the search space
in a NN search.  However, this  ability to find the relevant partitions in
space quickly degrades as data dimensionality increases.  This phenomenon is
the well-known ``curse of dimensionality'', which expresses the difficult in
efficiently partitioning the data or the space  as dimensionality
grows~\cite{Bohm:2001:SHS:502807.502809,Weber:1998:QAP:645924.671192}.

The approximate nearest neighbors (ANN) search was proposed to improve the
scalability of NN search in high-dimensional spaces for applications where
exact answers can be traded off for speed. A number of techniques and
algorithms for ANN search in high-dimensional spaces have been recently
proposed~\cite{Gionis:1999,Muja:2012,Valle:2008:HDI:1458082.1458181,5432202}.
FLANN~\cite{Muja:2012} builds a framework that dynamically selects the best
index, among algorithms such as randomized KD-trees~\cite{4587638},
hierarchical k-means~\cite{Nister:2006:SRV:1153171.1153548}, and
LSH~\cite{Gionis:1999}, for a given dataset.
Multicurves~\cite{Valle:2008:HDI:1458082.1458181} performs multiple projections
of subsets of the data objects dimensions to an 1-dimension space using
space-filling curves, and builds a sorted list for each projection. Its search
phase executes the same projections using the query object, and uses them to
retrieve the nearest points in the 1-dimension sorted lists as the candidates
for nearest neighbors. Finally, a ranking phase is executed to retrieve the
nearest elements from the set of candidates.  The Product Quantization based
ANN search~\cite{5432202} is another successful approach, which decomposes the
space into a Cartesian product of subspaces of lower dimensionality to further
quantize subspaces. The vector created from the quantized subspaces is then
used to estimate Euclidean distances.

Indexes based on \emph{locality sensitive hashing} (LSH)~\cite{Gionis:1999} 
are very popular and well recognized as
one of the most competitive techniques for similarity search in high-dimensional
spaces. We explain those techniques in more detail in Section~\ref{sec:lsh}.


\subsection{Parallel and Distributed ANN Search}

The ANN requirements for online CBMR systems are stringent and
include searching in very large datasets that grow fast as time passes, achieving high throughput, and providing low response times to end-users. These demands have motivated the development of ANN indexing methods
that make use of high performance
techniques~\cite{Stupar10,Bahmani:2012,search-CPU-GPU,Teodoro:2011:APA:2063576.2063651,teodoro-2013-vldb,4227958,10.1109/IPDPS.2012.45}.

The MapReduce based parallelizations of LSH~\cite{Stupar10,Bahmani:2012} are
the closest related works to ours. In the work of Stupar et.
al~\cite{Stupar10} the MapReduce formulation of LSH has: 1)~a \emph{map phase}
that independently visits buckets to which a query object is hashed to generate
a per bucket nearest neighbors set; and 2)~a \emph{reduce phase} to aggregate
results from all buckets visited during a query computation.  This LSH
implementation stores the buckets of points in a distributed file system (HDFS)
using a single file per bucket value. As reported by the authors, combinations
of LSH parameters may create a very large number of files (buckets) and
decreases the overall system performance. In addition, this implementation
stores data objects content in the bucket (files) for each hash table used,
instead of the object identifier (pointer) as in the original algorithm. As a
consequence, the entire dataset is replicated for each of the hash table used
by LSH. This level of data replication is prohibitive for large-scale datasets,
since LSH may require the use of several tables. Also, the high latency of data
accesses makes this solution impractical for online applications due to the
high query-processing times.

Bahmani et al.~\cite{Bahmani:2012} implemented another MapReduce-based variant
of the classic LSH algorithm that is referred to as Layered LSH. They have
implemented two versions of LSH using: 1)~Hadoop for file system based data
storage and 2)~Active DHT for in-memory data storage. They proposed theoretical
bounds for network traffic assuming that a single LSH hash table is used. This
assumption greatly simplifies the analysis and implementation of the algorithm,
but it may not be realistic because LSH typically achieves higher efficiency
with the use of several hash tables~\cite{Gionis:1999}. If multiple hash tables
are used, the theoretical propositions are not valid because the same data
object is indexed by multiple buckets from different hash tables and the data
partition would not be simple as in the case of a single table. Neither of the
MapReduce based parallelizations of LSH~\cite{Stupar10,Bahmani:2012} solves the
challenging problem of building a large-scale LSH-based searching index that
minimizes communication and avoids data replication, while preserving the
behavior of the sequential algorithm and providing low query response-times as
required in several online multimedia services.  As presented in
Section~\ref{sec:parallelization}, the parallelization strategy we propose in
the paper addresses all these limitations.


\section{Locality-Sensitive Hashing (LSH)}
\label{sec:lsh}

\begin{figure*}[htb!]
\begin{center}
\includegraphics[width=0.98\textwidth]{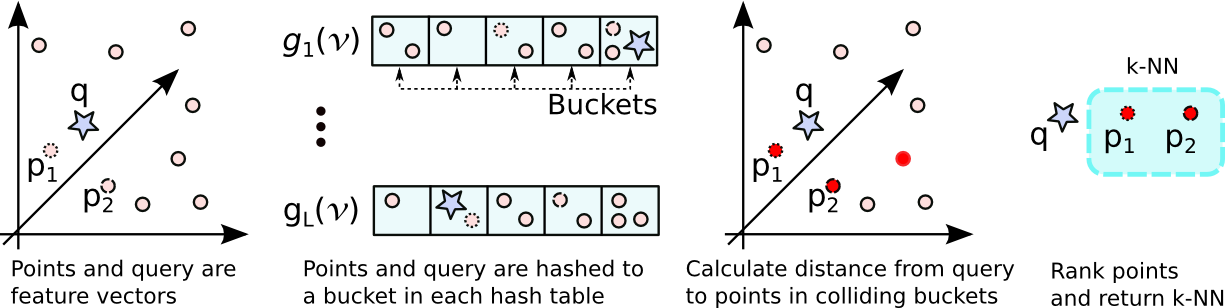}
\vspace*{-2ex}
\caption{The LSH execution scheme may be decomposed into the index building and
search phases. In the index building phase, each object in the reference dataset is mapped to a
single bucket in each of the L hash tables using hash function $g_j(v)$
(first two steps). During the search phase each query object ($q$) is also mapped
to data buckets using the same LSH functions, and those objects stored in
buckets to which $q$ was hashed are selected as candidates,  and then ranked according to their actual distance to $q$ in order to find
the $k$ nearest objects.
}
\vspace*{-2ex}
\label{fig:lsh-basic}
\end{center}
\end{figure*}

LSH employs locality-sensitive hash functions to assign objects to buckets via
quantization, such that similar objects are assigned together with high
probability. Indexing the dataset consists in hashing its objects to the
buckets.  The ANN search can then be carried out by (1)~finding the buckets to
which a query $q$ is hashed and selecting the objects in those buckets as
candidates, and (2)~ranking candidates according to their actual distance to
the query (See Figure~\ref{fig:lsh-basic}).



\subsection{Locality-Sensitive Hashing Functions}
\label{sec:lshbasic}

LSH relies on the existence on a family of locality-sensitive hashing
functions, from which  individual hashing functions can be sampled randomly.
The special property of the LSH function family that makes it useful for ANN
search is a guarantee that, as we keep sampling functions, they will tend to
hash together objects that are close, and hash apart objects that are distant.

More formally, assuming that $S$ is the domain of the objects, $U$ is the
domain of the hash keys, and $D$ is a distance function between objects:

\begin{mydef}
A function family $H = \{h: S \rightarrow U\}$ is called
$(r,cr,p_1,p_2)$-sensitive for $D$ if, for any $p,q \in S$:

  \begin{itemize}
      \item If $D(p,q) \leq r$ then $Pr_H[h(q)=h(p)] \geq p_1$,
      \item If $D(p,q) > cr$ then $Pr_H[h(q)=h(p)] \leq p_2$.
  \end{itemize}
\end{mydef}

In order to be useful for ANN search, we must have $c>1$ and $p_1 > p_2$, which
guarantees that objects within distance $r$ of the query ($q$) have higher
probability ($p_1$) of colliding with $q$ than those at distance greater than
$cr$ ($p_2$). 

The existence of locality-sensitive function families was proved by Indyk et
al.~\cite{Indyk1998}, in the seminal paper that also introduced LSH indexing.
Since then, other families have been introduced (each family works for a
specific combination of data domain and distance function). Of particular
interest for practical applications are the  $p$-stable family~\cite{Datar2004}
that exploits the stability of the Gaussian distribution under the
$\ell_2$-norm to hash data in the Euclidean spaces. 

An individual hash function from the $p$-stable family is defined as:

\begin{equation}
 h_{a,b}( v ) =  \left\lfloor \frac{  a \cdot v  + b } {w} \right\rfloor
\end{equation}

\noindent
where $ a \in \mathbb{R}^d $ is a random Gaussian vector sampled from $\mathcal{N}(0,I)$, and
$b$ is a random offset scalar sampled from $\operatorname{unif}\left(0, w \right)$. Sampling the values of $a$ and $b$ from their distributions consists,
effectively, in sampling an individual hash function from the family.  The parameter $w$ is fixed for the entire family, and acts as a quantization width. Applying $h_{a,b}$
to a vector or object $v$ corresponds to the composition of a projection
to a random direction (with random offset) and a quantization
given by a constant scaling and the floor operation.

\subsection{LSH Indexing}
\label{sec:lshindexing}

Each individual hash function is locality-sensitive, but not extremely so ---  there is no particular guarantee making $p_1$ close to 1 or $p_2$ close to 0, just $p_1 > p_2$. In order to make locality-sensitiveness actually useful,
LSH employs a function family $G$, created by concatenating $M$
hash functions from $H$, i.e., ${G = g : S \rightarrow U^M}$, such that
each $g \in G$ has the following form: $g(v)=(h_1(v),\ldots,h_M(v))$, where
$h_i \in H$ for $1 \le i \le M$. 

As we concatenate multiple $h_i\in H$ functions into a single $g$, the
probability of a \emph{false positive} (distant vectors hashed together)
decreases exponentially. Unfortunately, so increases the probability of a
\emph{false negative} (close vectors hashed apart). In order to solve the
latter problem, multiple  functions $g_j \in G$ ($1 \le j \le L)$ are employed
to build $L$ independent hash tables, in the hope that the good answers will be
found in at least one of them. Indyk et al.~\cite{Indyk1998} prove formally
that $M$ and $L$ can be chosen such that for any query, there is a high
probability of finding the right answer, without having to examine too many
vectors (false positives). The informal argument that we offer here is that
false positives have probability $p_2^M$, which clearly falls very fast w.r.t.
$M$. As $L$ grows, both false positives and true positives tend to grow, but the former at a rate related to $p_2$ and the latter at rate
related to $p_1$ (which is $>p_2$), so it is intuitive that the true positives
will grow much faster. Therefore we can find a suitable combination of $L$ and
$M$, with high true positives and fairly low false positives.  

As we have explained in the beginning of the section, LSH may be conceptually
decomposed into the phases of index building, and of searching. It is clear now
that several hash tables must be employed both during index and search, one for
each hashing functions $g_j \in G$. In the standard implementations, a limit is
imposed on the maximum number of candidates to be retrieved (usually $2L$ or
$3L$) in order to limit the worst case of distances to be computed per query.
Also, there is no replication of the dataset: the multiple hash tables only
keep a reference to another structure where objects are effectively stored
(e.g. a position in an array, or a pointer), so after visiting the buckets, the
method has to retrieve the actual data. This is usually fairly cheap for
shared-memory environments with uniform access cost. But when access cost is
not uniform (disk storage, distributed memory), the implementation of LSH
becomes extremely challenging and has been an open problem.



\subsection{Improvements on the basic LSH Indexing}
\label{sec:advlsh}

The popularity of  LSH  motivated much research 
focused on improving its performance, and reducing its memory footprint. 
Memory usage is one of the main limitations of the standard LSH,
since it has to employ several hash tables in order to reduce the probability of false negatives,
a strategy that favors very large values of $L$ (from a few dozens to a few hundreds).  
Even though the hash tables store just references, for such large values of $L$ memory usage quickly becomes a concern.

Entropy-based nearest neighbor search~\cite{Panigrahy:2006} is one of the
first efforts to address this deficiency. In this approach fewer
hash tables are used, but multiple buckets are accessed in each hash table. The
buckets visited in each table are chosen by hashing objects randomly selected in
the neighborhood of the query object. Because those objects are close to the
query object, they are also expected to hash to buckets that have objects similar to the query in the original $d$-dimensional space. 

Multi-probe LSH~\cite{Lv:2007} further extended the entropy based approach with
the introduction of a careful methodology for selecting multiple buckets to be
accessed in each hash table. Instead of using random objects in the query
neighborhood, it estimates the likelihood of a bucket containing good answers
from its distance to the query, thus directly deriving the buckets to be
visited. This approach typically results, for the same recall, in less bucket
accesses per hash table as compared to entropy-base LSH.  An extension of this
work interprets the likelihoods as full-fledged probabilities, by learning the
priors from a sample of the datasets, in a scheme named A~Posteriori
LSH~\cite{Joly2008a}. A Posteriori potentially reduces the number of buckets to
visit, but the computation of the list of buckets to visit becomes
significantly more expensive. 

Several works also studied different locality-sensitive hash
functions~\cite{Pauleve2010}; support for a diversified set of similarity
functions~\cite{Kulis:2012:KLH:2197080.2197253} and improved approach for
calculating distance~\cite{Dasgupta:2011:FLH:2020408.2020578}; the use of
query-adaptive LSH with bucket-filtering based on the query position relative
to the quantitized frontier~\cite{Jegou2008}; and the automatic tuning of the
LSH parameters~\cite{Bawa2005,DBLP:journals/pieee/SlaneyLH12} such as the
number of hash tables to be used.

\section{Distributed Memory LSH Indexing}
\label{sec:parallelization}

The parallelization strategy we employ is based on the dataflow programming
paradigm~\cite{ARPACI-99,beynon01datacutter,10.1109/ICPP.2008.72}.
Dataflow applications are typically represented as a set of computing
\emph{stages}, which are connected to each other using directed \emph{streams}.
The rest of this section is organized as follows: Section~\ref{sec:parStrategy}
presents an overview of our parallelization strategy focusing on the basic LSH
method; Section~\ref{sec:intra} discusses the implementation of the
intra-stage parallelization; Section~\ref{sec:mapping} details different mapping
strategies we have evaluated to partition the application state in a
distributed environment; and Section~\ref{sec:pmprobe} presents extensions in
our parallel algorithm to support the execution of the multi-probe based LSH
scheme.

\subsection{Parallelization Strategy}
\label{sec:parStrategy}

\begin{figure*}[htb!]
\begin{center}
\includegraphics[width=0.94\textwidth]{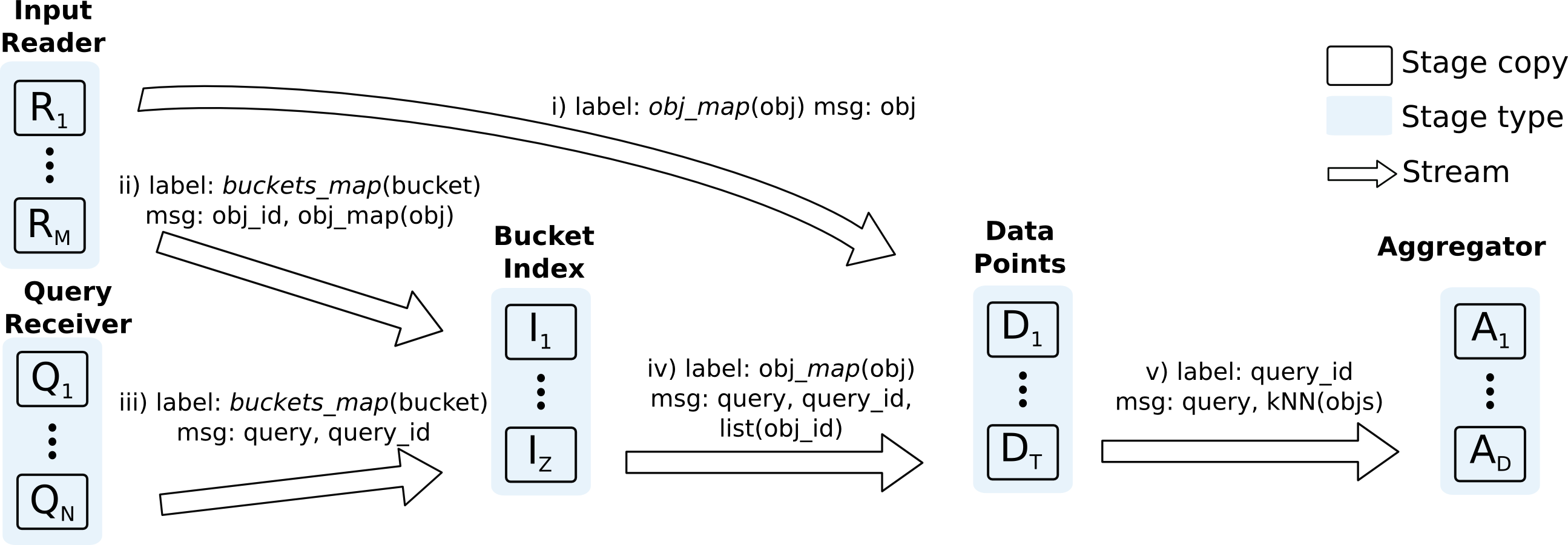}
\vspace*{-2ex}
\caption{LSH decomposition into the dataflow programming paradigm. The index
building phase partitions the input dataset among the DP copies without data
replication (message~i) and the BI stage stores buckets of objects identifiers
(message~ii). During the search phase the buckets in which the query is hashed are calculated in the
QR that communicate with the BI stage to visit those buckets of interest.  Further, BI
copies send messages to the DPs copies that store at least one point of interest found
on the buckets visited. DPs then calculate the local NN using those points it stores.
Finally, DP copies will send the local NN results for reduction with the AG
stage.}
\vspace*{-2ex}
\label{fig:parallel}
\end{center}
\end{figure*}

Our parallelization decomposes LSH into five computing stages organized in two
conceptual pipelines, which execute the index building and the search phases of
the application (Figure~\ref{fig:parallel}). All stages may be replicated in
the computing environment to create as many copies as necessary. Additionally,
the streams connecting the application stages implement a special type of
communication policy referred here as \emph{labeled-stream}.  Messages sent
through a labeled-stream have an associated label or tag, which provides an
affordable scheme to map message tags to specific copies of the receiver stage
in a stream. This tag-to-receiver mapping is computed using a hash function
called by the labeled-stream for each message sent.  We rely on this
communication policy to partition the input dataset and to perform parallel
reduction of partial results computed during a query execution. The data
communication streams and processes management are built on top of Message
Passing Interface (MPI).

\emph{The index building phase} of the application, which includes the Input
Reader (IR), Bucket Index (BI), and Data Points (DP) stages, is responsible
for reading input data objects and building the distributed LSH indices that
are managed by the BI and DP stages.  In this phase, the input data objects are
read in parallel using multiple IR stage copies and are sent (1)~to be stored
into the DP stage (message~i) and (2)~to be indexed by BI stage (message~ii).
First, each object read is mapped to a specific DP copy, meaning that there is
no replication of input data objects. The mapping of objects to DPs is carried
out using the data distribution function \emph{obj\_map} (labeled-stream
mapping function), which calculates the specific copy of the DP stage that
should store an object as it is sent through the stream connecting IR and DP.
This mapping may be calculated based on some characteristic of the object,
e.g., its location on the space or simply on the object identifier. Further,
the pair $<$object identifier, DP copy in which it is stored$>$ is sent to each
BI copy that holds buckets to which the object was hashed.  The distribution of
buckets among BI stage copies is carried out using another mapping function:
\emph{bucket\_map}, which is calculated based on the bucket value/key. Again,
there is no replication of buckets among BIs and each bucket value is stored
into a single BI copy.  

For sake of simplicity, we assume in this section that both labeled-stream
mapping functions (\emph{obj\_map} and \emph{bucket\_map}) are a mod operation
that is calculated using either the obj\_id or the bucket value (tags) and the
number of copies of the receiver stage, e.g.  $obj\_id\ mod\ T$ for the
\emph{obj\_map} and DP stage. However, we have evaluated different classes of
mapping functions as detailed in Section~\ref{sec:mapping}. The labels and the
body of messages in the communication from IR to DP and IR to BI are presented
in Figure~\ref{fig:parallel}. Additionally, it is important to highlight that
our labeled-stream implementation employs buffering and aggregation of messages
to maximize network performance. Therefore, when messages are dispatched using
the stream interface they may be first copied to a buffer for aggregation with
other messages before they are sent over the network. This strategy intends to
improve performance in the communication among stages such as IR and DP, for
instance, in which sending a single small message (object reference) would
result in under-utilization of the network and high overheads. 

After the index building phase has finished, the buckets created by the $L$ hash
tables are distributed through the BI stage copies. Each of the buckets, as
discussed, stores only the identifier of the objects and the copy of DP stage in
which they are located.  

\emph{The search phase} of the parallel LSH uses four stages, two of them
shared with the index building phase: Query Receiver (QR), Bucket Index (BI),
Data Points (DP), and Aggregator (AG).  The QR stage reads the query objects
and calculates the bucket values in which the query is hashed for the $L$ hash
tables used. Each bucket value computed for a query is mapped to a BI copy
using the \emph{bucket\_map} function. The query is then sent to those BI stage
copies that store at least one bucket of interest (message~iii). Each BI copy
to receive a query message visits the buckets of interest, retrieves the
identifier of the objects stored on those buckets, aggregates all object to be
sent to the same DP copy (list(obj\_id)), and sends a single message to each DP
stage that stores at least one of the retrieved objects (message~iv). For each
message received by a DP copy, it calculates the distance from the query to the
objects of interest, selects the k-nearest neighbors objects to the query, and
sends those local NN objects to the AG stage. Finally, the AG stage receives
the message containing the DPs local NN objects from all DPs involved in that
query computation and performs a reduction operation to compute the global NN
objects. As presented in Figure~\ref{fig:parallel} (message~v), DP copies use
the query\_id as a label to the message, what guarantees that the same AG copy
will process all messages related to a specific query.  As a consequence,
multiple AG copies may be created to execute different queries in parallel.
Although we have presented the index building and the search as sequential
phases for sake of simplicity, their execution may overlap.

The parallelization approach we have proposed exploits task, pipeline,
replicated and intra-stage parallelism. Task parallelism results of concurrent
execution of IR and QR that allows for indexing and searching phases to
overlap, e.g.  during an update of the index. Pipeline parallelism occurs as
the search stages, for instance, execute different queries in parallel in a
pipeline fashion. Replicated parallelism is available in all stages of the
application, which may have an arbitrary number of copies. Finally, intra-stage
parallelism results of the application's ability to use multiple cores within a
stage copy as detailed in Section~\ref{sec:intra}.

\subsection{Intra-Stage Parallelization}
\label{sec:intra}

The intra-stage parallelization refers to a stage ability to execute tasks in
parallel using multiple computing cores available in a node. This level of
parallelism is important to fully take advantage of current machines, which are
typically built as multi-socket, multi-core systems. One of the main advantages
in employing intra-stage parallelization, as compared to creating one stage
copy per computing core as in classic MPI-based parallelizations, refers to the
possibility of sharing the same memory space among computing cores used in a
stage copy. In stateful applications such as LSH, a smaller number of state
partitions may be created and, as a consequence, a reduced number of messages
needs to be exchanged during the computation of a query.  

The intra-stage parallelism in our application is implemented using POSIX
Threads, and is employed to compute messages arriving at
the Buckets Index and the Data Points stages in parallel. These stages have
been selected as target for intra-stage parallelization because (i)~they are
application's most compute intensive stages and (ii)~they store the
application index and input dataset (state) and, therefore, this strategy leads
to a reduced network traffic and better application scalability.  Messages
arriving at both stages are independently processed in an embarrassing parallel
fashion using all the computing cores available in a node. As discussed, this
allows for a single copy of these stages to be created in each machine in which
they are executed.

\subsection{Data Partition Strategies}
\label{sec:mapping}

This section considers the partition of the data with our parallel LSH. As
presented in the previous section, the $buckets\_map$ and $obj\_map$ functions
are responsible for mapping, respectively, buckets and objects (data points)
from the input dataset into one of the BI and DP stage copies. The adequate
partition of the dataset is of major importance for the performance, because it
directly impacts the number of BI and DP copies that need to be consulted (and
messages sent) to process a given query. 

Therefore, the partition of the application state should be performed in way to
minimize the number of messages exchanged. This resembles the original problem
addressed by indexing algorithms, which are in essence trying to partition a
high-dimensional space in a way that can be used to efficiently search on that
space. To address this challenge, we have studied the impact different mapping
functions to the application performance. The functions evaluated include those
with locality preserving properties, which would preferably map data points
close in the space to the same machine (partition).

The mapping functions analyzed are: (1)~a \emph{mod} operation that is
calculated based on buckets and data points identifiers and does not preserve
data locality (presented in Section~\ref{sec:parStrategy}); (2)~a
\emph{space-filling} curve that is a locality preserving fractal curve
introduce by Peano and Hilbert~\cite{Valle:2008:HDI:1458082.1458181}. Data points in a
d-dimensional space are mapped to a position in the curve, and that position is
used as an indicative of how close data points are in the space.  We have
specifically used the Z-order curve, which is calculate with a bit shuffle and
has same properties such as other curves more complex to compute (e.g., Hilbert
curve); (3)~a \emph{LSH} function that, as discussed, tend to map objects close
in the space to the same hash value. This mapping has used an instance of the
$g(v)$ function (Section~\ref{sec:lshindexing}) different from those used to
build the index.

\subsection{Support to Multi-Probe LSH}
\label{sec:pmprobe}

As previously discussed in Section~\ref{sec:advlsh}, LSH may need to use
several hash tables to attain the desired search quality, which could lead to
high memory usage. Therefore, in our parallel version, we have also developed
support for executing the Multi-probe LSH~\cite{Lv:2007} to perform multiple
probes ($T$) in each hash table create, such that the desired quality may be
reached using a smaller number of hash tables.

In this approach, probes generated for the original query vector have an unique
identifier (bucket) in each hash table, and the execution of each of them is
treated similarly to a independent query in the search phase pipeline, until
the AG stage. During the AG phase, however, those probes generated from the
same original query need to be aggregated to create a global unique query
answer. In our implementation, we developed an extra message aggregation level
in order to pack messages from different probes related to the same query that
are routed to the same stage copies. As presented in the experimental
evaluation, this approach is very effective and directly impacts in a positive
way to the application execution time.

\section{Experimental Results}
\label{sec:results}

%

\subsection{Experimental Setup}
\label{sec:setup}

The experimental evaluation was performed using a distributed-memory machine
with 60 nodes.  Each computation node is equipped with a dual-socket Intel E5
2.60~GHz Sandy Bridge processor with a total of 16 CPU cores, 32~GB of DDR3 RAM,
and runs Linux OS kernel version 2.6.32.  The nodes are interconnected through
a FDR Infiniband switch.

\begin{table}[h]
\caption{Datasets of 128-dimensional SIFT feature vectors used in our evaluation.}
\vspace{-2mm}
\begin{center}
{\small
\begin{tabular}{ l l l }
\hline
\textbf{Name} 	& \textbf{Reference set size} 	& \textbf{Query set ($Q$) size} 	\\ \hline
Yahoo 		& 130 million 				& 233,852 			\\
BIGANN 		& 1 billion 				& 10,000 			\\ \hline
\end{tabular}
}
\end{center}
\vspace{-4mm}
\label{tab:datasets}
\end{table}

For the evaluation, we have employed the single most cited feature
vectors employed for images, SIFT~\cite{Lowe:2004:DIF:993451.996342}, which have 128 dimensions. Two datasets were used (see
Table~\ref{tab:datasets}). The \emph{Yahoo}~\cite{teodoro-2013-vldb} dataset
contains about 130 millions SIFT vectors that were extracted from 225 foreground images from our personal collections, and 233,852 background Web images, employed to confound the method. It contains 187,839 query feature vectors created from strong geometric and photometric distortions of the foreground images. The \emph{BIGANN}~\cite{5946540} is to the best of our knowledge the largest
dataset for multimedia retrieval publicly available.  It contains 1 billion
128-dimensional SIFT feature vectors computed from 1 million Web images and a
query feature vector set with 10,000 vectors. Both datasets have the
ground-truth calculated for the query, which is used to evaluate the search
quality.

The search quality was  the \emph{recall}~\cite{Lv:2007}, which measures the fraction of the true $k$ nearest neighbors that where effectively retrieved by the method.

%
%

\subsection{Assessing the Distributed LSH Scalability}

This set of experiments focuses on evaluating the parallel multi-probe LSH
scalability. Our analysis employs a scale-up (weak scaling) experiment in which
the reference dataset and the number of computing cores used increase
proportionally.  A scale-up evaluation was selected because we expect to obtain
an abundant volume of data for indexing, which would only fit in a distributed
system. The Yahoo dataset is used along with a round-robin data partition
strategy. Further, the LSH parameters employed are $L = 6$ and $M = 32$.  A
detailed evaluation of the parameters impact to the execution times and search
quality is presented in the following sections. Additionally, in these
experiments a single CPU core is allocated for the AG stage, while the number
of computing cores used by the BI and DP stages increase. The ratio of
computing cores used by BI:DP stages is 1:4, meaning that for each CPU core
allocated to the BI stage 4 CPU cores are assigned to the DP stage. In all
experiments in this paper we have requested the algorithm to retrieve the
10-nearest neighbors ($k=10$).

\begin{figure}[htb!]
\begin{center}
\includegraphics[width=0.49\textwidth]{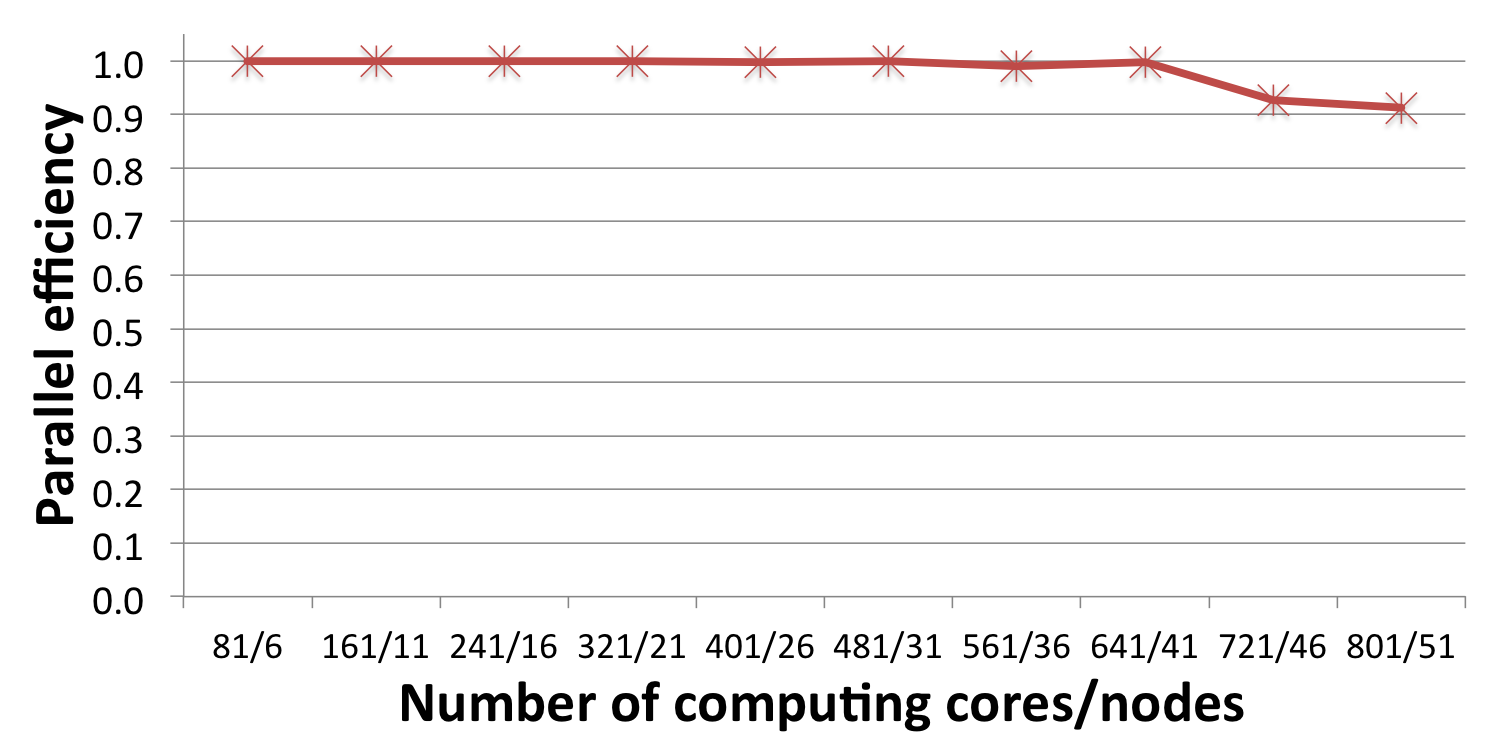}
\vspace{-4ex}
\caption{Efficiency of the multi-probe LSH parallelization using a weak scaling
evaluation in which the number of nodes used and the reference dataset size increase
proportionally.}
\vspace{-2ex}
\label{fig:scalability} 
\end{center} 
\end{figure}

The efficiency of the parallel multi-probe LSH as the number of CPU cores and
nodes used increase is presented in Figure~\ref{fig:scalability}. As shown, the
application achieves a very good parallel efficiency, e.g. about 0.9 is
attained when 801 computing cores are used (10 nodes for BI and 40 nodes for
DP).  The high efficiency attained is a result of (i)~the application
asynchronous design that decouples communication from computation tasks and of
(ii)~the intra-stage parallelization that allows for a single multi-threaded
copy of the DP stage to be instantiated per computing node. As a consequence, a
smaller number of partitions of the reference dataset are created, which
reduces the number of messages exchanged by the parallel version (using 51
nodes) in more than 6$\times$ as compared to an application version that
instantiates a single process per CPU computing core.

\subsection{Evaluating the Multi-Probe LSH Performance}

This section evaluates the distributed multi-probe LSH with respect to the
trade-offs between the search quality and execution time as the number of
probes ($T$) per table is varied.  The BIGANN dataset (1 billion 128-dimensional
SIFT vectors and a query set with 10,000 descriptors) is employed, and the LSH
parameters chosen after tuning are $L = 6$ and $M = 32$. All experiments are
executed using 801~CPU cores/51 nodes.


\begin{figure}[htb!]
\begin{center}
\includegraphics[width=0.49\textwidth]{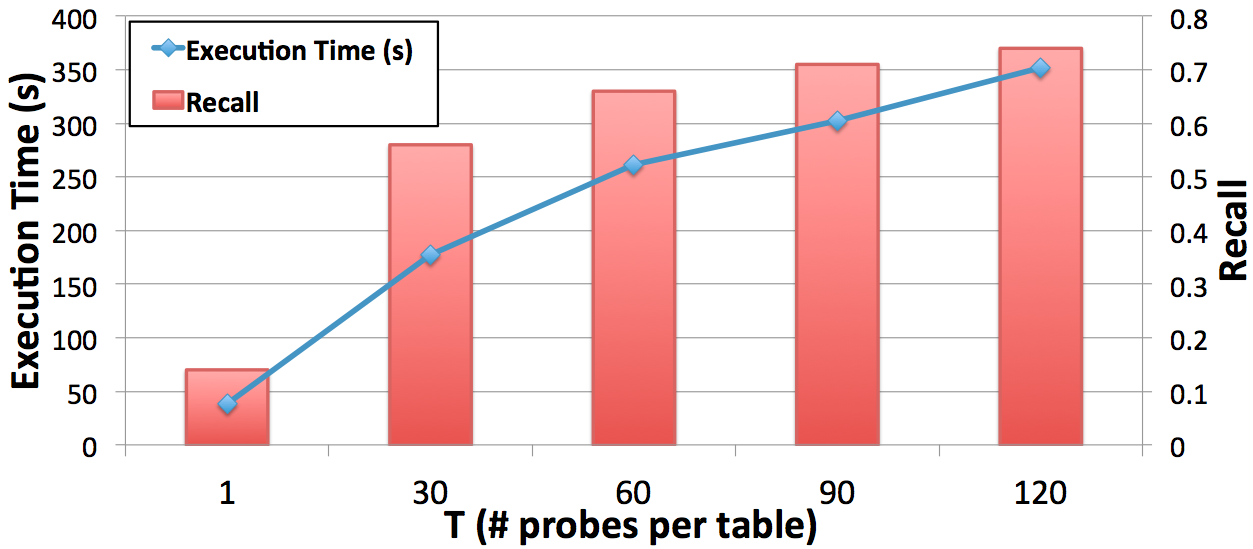}
\vspace{-4ex}
\caption{Multi-probe LSH evaluation in large-scale runs using 801~CPU cores/51 nodes and the BIGANN dataset (1 billion vectors): the trade-off between execution times and search quality as the number of probes (buckets visited) per table increase.}
\vspace{-2ex}
\label{fig:mp}
\end{center}
\end{figure}

The compromise between execution time and search quality (recall) according to
$T$ is presented in Figure~\ref{fig:mp}. As expected, the recall is improved with
the increasing of $T$, but the execution time grows sublinearly with respect to
the number probes ($T$) performed. For instance, the variation in $T$ from 60 to
120 results into an execution time increase of only 1.35$\times$, although it
was expected that the ratio of the changes in $T$ and execution time would be
roughly the same. This is a very good property of our parallel LSH that
benefits the use of larger values of $T$ to improve search quality with reduced
impact to execution times.

\begin{table}[h]
\caption{Profile of network communication as $T$ is varied: the volume of
data and number of messages exchanged.}
\vspace{-4mm}
\begin{center}
\begin{tabular}{ l l l }
\hline
\textbf{T} 	& \textbf{Volume of Data(GB)} 	& \textbf{\# of Messages. ($\times 10^6$)} \\ \hline
1		& 2.28 				& 1.83 					\\
30 		& 35.73 			& 49.43 				\\ 
60 		& 59.46 			& 94.23 				\\ 
90 		& 79.31 			& 136.55 				\\ 
120 		& 96.82 			& 177.08 				\\ \hline
\end{tabular}
\end{center}
\vspace{-4mm}
\label{tab:mp-msg}
\end{table}

The lower increasing ratio in execution time as consequence of variations in
$T$ is a result of: (1)~the aggregation of all messages related to bucket
visits ($T$ per table) that are routed to the same stage copy; and,
(2)~elimination of duplicated distance calculations that occur when the same
data point is retrieve multiple times from different hash tables. The
probability of such duplications is higher as $T$ increases. As presented in
Table~\ref{tab:mp-msg}, the volume of data~(GB) and the number of messages
exchanged by the application increase, respectively 1.22$\times$ and
1.29$\times$ as $T$ is varied from 60 to 120.

%
%

\subsection{The Impact of LSH Parameters to Performance}

This section evaluates the impact of the LSH parameters ($M$ and $L$) to the
search quality and execution times using the BIGANN dataset.  The value of $M$
(hash functions used per hash table) affects the selectivity of the index. High
values of $M$ will lead to reduced probability of mapping far away objects to
the same bucket, but too high values may also result in the mapping of close
object to different buckets (with smaller probability). The $L$ parameter
specifies the number of hash tables used and, as a consequence, visited during
the search. A tuning phase is typically employed to determine the optimal
values of $M$ and $L$ for a given dataset, since the interactions among these
parameters are complex.

In order reduce the combination of parameters evaluated in large-scale
experiments in this section, we have first tuned $M$ and $L$ using the
sequential version of the multi-probe LSH and a smaller partition of the
dataset.  Further, we varied $M$ in the neighborhood of the optimal value for
the sequential algorithm ($M=30$) and increased the value of $L$ until we reach
the amount of memory available for the BI stage in our configuration.

\begin{table}[h]
\caption{Impact of varying the number of hash functions ($M$) used by each LSH hash table.}
\vspace{-4mm}
\begin{center}
\begin{tabular}{ l l l }
\hline
 	        & \textbf{Execution Time (s)} 	& \textbf{Recall} \\ \hline
M=28       	& 3,463.47  & 0.8  \\
M=30      	& 264.84  & 0.73  \\
M=32       	& 261.73  & 0.66  \\ \hline
\end{tabular}
\end{center}
\vspace{-4mm}
\label{tab:impM}
\end{table}


The impact of varying $M$, for fixed values of $T=30$ and $L=6$, is presented in
Table~\ref{tab:impM}. As expected, the recall slowly decreases with the
increasing of $M$ due to the higher selectivity of the hash table. The search
time, however, quickly decreases (about an order of magnitude) for values of $M$
higher than 28. This shows that a high selectivity is important for the
performance of the algorithm, and that the best value of $M$ for the distributed
version is also in the neighborhood of 30.


\begin{figure}[htb!]
\begin{center}
\includegraphics[width=0.49\textwidth]{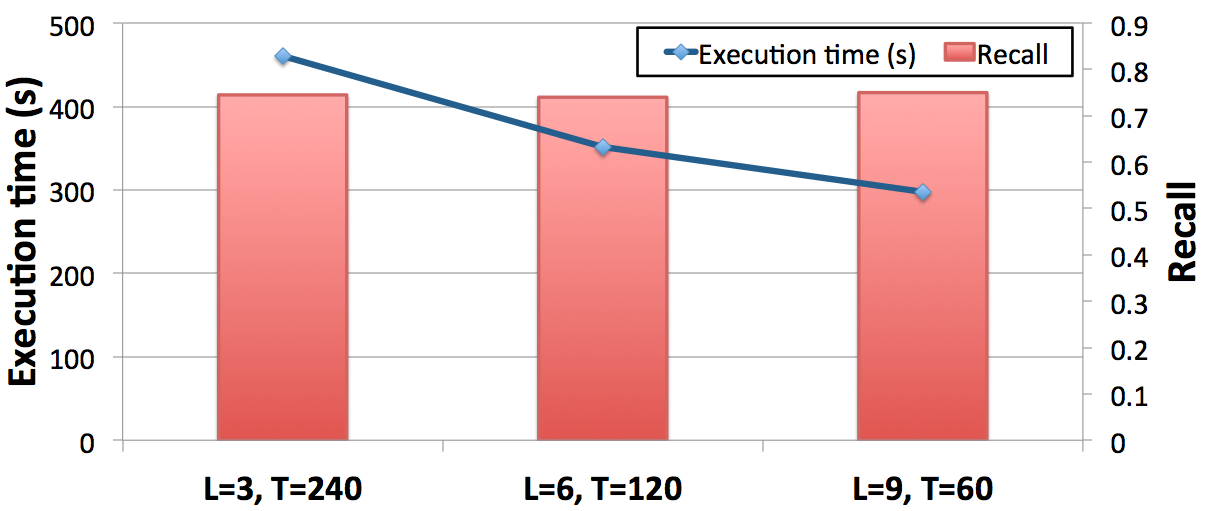}
\vspace{-4ex}
\caption{Impact of varying the number of hash tables (L) and T used by LSH for the same level of search quality.}
\vspace{-2ex}
\label{fig:impL}
\end{center}
\end{figure}

Finally, Figure~\ref{fig:impL} shows the performance of the distributed
multi-probe LSH as the number of hash tables ($L$) employed is varied. In these
experiments, we have increased the value of $T$ for each value of $L$ until the
same level of recall is achieved (about 0.74). As shown, the largest number of
hash tables lead to better execution times for similar recall values. However,
the increasing in the number of hash tables also result into higher memory
demands, which limit the number of hash tables that may be used by the
application.

\subsection{Effectiveness Of Data Partition Strategies}

This section presents an evaluation of the data partitioning using the three
mapping functions mentioned in Section~\ref{sec:mapping}:~\emph{mod} operation,
Z-order curve, and \emph{LSH} function. The BIGANN dataset is used for all
experiments, and the LSH parameters chosen are $L = 6$, $M = 32$, and $T = 60$.
All experiments used the full environment with 801 CPU cores/51 nodes (10 nodes
for BI and 40 nodes for DP).

\begin{figure}[htb!]
\begin{center}
       \includegraphics[width=0.47\textwidth]{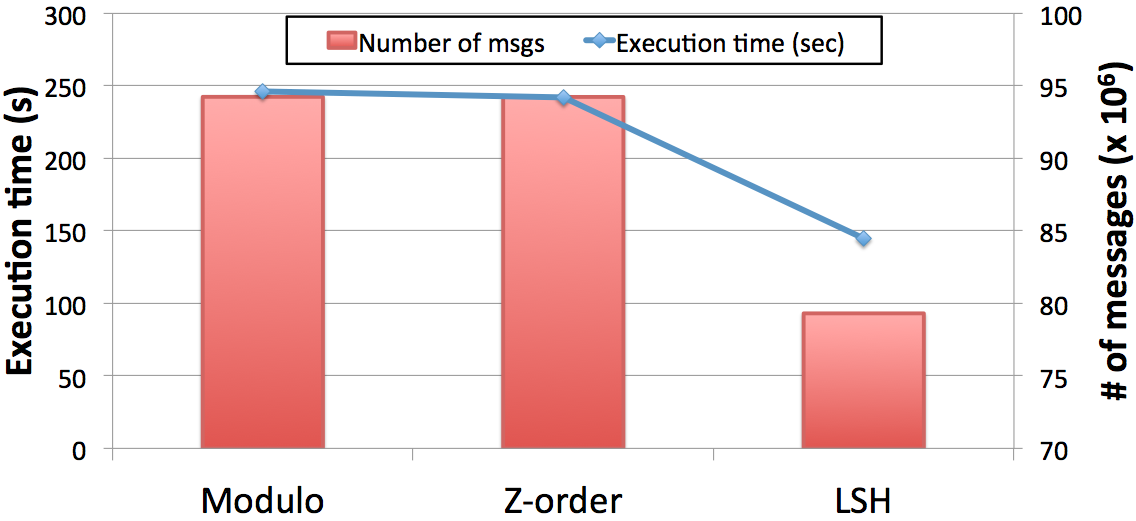}
\vspace*{-2ex}
\caption{Evaluation of Data Partition Strategies.}
\vspace*{-2ex}
\label{fig:data-partition-study}
\end{center}
\end{figure}

The execution time and the number of messages exchanged for each mapping
function are presented in Figure~\ref{fig:data-partition-study}.  As may be
noticed, the execution time for \emph{mod} (246 seconds) and Z-order (242
seconds) are similar. The use of the \emph{LSH} mapping, however, improved the
performance on top of the other strategies in at least 1.68$\times$. The
differences in execution time are derived from the better LSH mapping, which
also results into lower number of messages exchanged by the application.

We have also evaluate the level of load imbalance regarding the number of
objects mapped to each DP copy for each mapping strategy. The load imbalance is
defined here as difference from the actual number of data objects assigned to
DP copies to the average. For the \emph{mod} mapping, as expected, there is no
load imbalance, whereas it was 0.01\% and 1.80\%, respectively, for the
\emph{Z-order} and \emph{LSH}. This shows that LSH performs an efficient data
partition without incurring in significant load imbalance.

\section{Conclusions and Future Work}
\label{sec:conclusions}

The similarity search is one of the most costly phases of query processing in
content-based multimedia retrieval applications and, as a consequence, is
critical for the success of these services.  This paper presents an efficient
and scalable LSH based similarity search index for large-scale and
high-dimensional multimedia datasets. Our approach addresses the challenges of
the LSH indexing parallelization by avoiding  data replication, using
hierarchical parallelization to take advantage of all CPU cores in a node,
studying multiple data partition strategies, overlapping communication and
computation, and using a multi-probing strategy to reduce the memory footprint.
We evaluated our implementation using a dataset with 1 billion 128-dimensional
SIFT feature vectors extracted from images collected in the Web, which is
to the best of our knowledge the largest publicly available dataset. 

The parallel LSH has attained about 90\% of efficiency in a parallel execution
using up to 801 CPU cores. In addition, the use of multi-probe with our
optimization to group messages resulted into very good trade-offs between
quality and search time, since the application execution time grows sublinearly
with the increasing of the number of probes ($T$) performed.  We have also
shown that there is a strong impact of the number of hash tables ($L$) and the
number of hash functions per hash table ($M$) to system's performance.
Finally, we presented an evaluation of three data partition strategies,
\emph{mod} operation, Z-order curve, and \emph{LSH} function, in which the
latter attained a performance improvement of at least 1.68$\times$ on top of
the others.

It is perhaps worth to emphasize that LSH is one of the most important
high-dimensional indexes proposed in the literature, but until now, its
performance has been confined to shared-memory uniform-access systems, since
previous attempts to parallelize it have stumbled upon the challenge of
adapting its lack of referential locality to scalable, distributed memory
architectures. By addressing those challenges, our technique brings one of the
most successful sequential high-dimensional indexes to the operational scenario
of large-scale online content-based multimedia services. As future work, we
plan to use accelerators, such as GPUs and Intel Xeon Phi, to perform
coordinated LSH execution using CPUs and
accelerators~\cite{DBLP:conf/ipps/TeodoroPKKCPKS13,DBLP:journals/pc/TeodoroPKKCS13,cluster09george,Teodoro:2012:Cluster}.
We also want to test our implementation in other areas, such as biology, in
order to find similarities in genomic databases.

\bibliographystyle{IEEEtran}
\begin{scriptsize}
\bibliography{george}  
\end{scriptsize}

\end{document}